\documentclass{aa}  

\usepackage{graphics}
\usepackage[abs]{overpic}
\usepackage{txfonts}
\usepackage{multirow}
\usepackage{longtable}
\begin{document}

   \title{Impact of micro-telluric lines on precise radial velocities and its correction}

   %\subtitle{I. Overviewing the $\kappa$-mechanism}

   \author{D. Cunha \inst{1,2}
	  \and
          N. C. Santos \inst{1,2}
          \and
          P. Figueira \inst{1}
          \and
          A. Santerne \inst{1}
	  \and
	  J. L. Bertaux \inst{3}
          \and
	  C. Lovis \inst{4}
%\fnmsep\thanks{Just to show the usage
%          of the elements in the author field}
          }

   \institute{Centro de Astrof\'{i}sica da Universidade do Porto, Rua das Estrelas, 4150-762 Porto, Portugal\\
              \email{Diana.Cunha@astro.up.pt}
         \and
             Departamento de F\'{i}sica e Astronomia, Faculdade de Ci\^encias, Universidade do Porto, Portugal
	 \and 
	     Universit\'e Versailles Saint-Quentin; Sorbonne Universit\'e, UPMC Univ. Paris 06; CNRS/INSU, LATMOS-IPSL, 11 Boulevard d’Alembert, 78280 Guyancourt, France
         \and 
	     Observatoire Astronomique de l'Universit\'e de Gen\`eve,
            51 Ch. des Maillettes, - Sauverny - CH1290, Versoix, Suisse}

   \date{Received \today   Accepted Month xx, 2014}

% \abstract{}{}{}{}{} 
% 5 {} token are mandatory
 
  \abstract
  {In the near future, new instruments such as ESPRESSO will arrive, allowing us to reach a precision in radial-velocity measurements on the order of 10 $\mathrm{cm}\,\mathrm{s}^{-1}$.
At this level of precision, several noise sources that until now have been outweighed by photon noise will start to contribute significantly to the error budget. 
The telluric lines that are not neglected by the masks for the radial velocity computation, here called micro-telluric lines, are one such noise source.
}
  {
 In this work we investigate the impact of micro-telluric lines in the radial velocities calculations. 
We also investigate how to correct the effect of these atmospheric lines on radial velocities.}
   {The work presented here follows  two parallel lines. 
First, we calculated the impact of the micro-telluric lines by multiplying a synthetic solar-like stellar spectrum by synthetic atmospheric 
spectra and evaluated the effect created by the presence of the telluric lines. 
Then, we divided HARPS spectra by synthetic atmospheric spectra to correct for its presence on real data and calculated the radial velocity on the corrected spectra.
 When doing so, one considers two atmospheric models for the synthetic atmospheric spectra: the LBLRTM and TAPAS.}
   {We find that the micro-telluric lines can induce an impact on the radial velocities calculation that can already
 be close to the current precision achieved with HARPS, 
and so its effect should not be neglected, especially for future instruments such as ESPRESSO. 
Moreover, we find that the micro-telluric lines' impact depends on factors, such as the radial velocity of the star, airmass, relative humidity, 
and the barycentric Earth radial velocity projected along the line of sight at the time of the observation.}
   {}

   \maketitle
%
%________________________________________________________________

\section{Introduction}
{ Since the discovery in 1995 of the first planet orbiting another star than the Sun by \citet{Mayor_Queloz_1995}, more than 
1000 exoplanets have been discovered\footnote{reported in exoplanet.eu in Feb 25, 2014}. From these, more than  half were detected using the radial velocity (RV) technique. 
The RV method consists in  measuring  the Doppler shifts of the stellar spectra 
caused by the back and forth movement of the star around the center of mass of the planetary system. 
The current ace in exoplanet discovery through the RV method is HARPS, a fiber-fed, cross-dispersed
echelle spectrograph installed on the 3.6-m telescope at  La Silla Observatory, which routinely achieves a precision 
 better than $1\,\textrm{ms}^{-1}$ \citep{Mayor_et_al_2003} and down to $50 - 60\,\textrm{ms}^{-1}$ \citep{Dumusque_2012}. 
{The RV calculation from HARPS observations is done using the cross-correlation function (CCF) technique \citep{Baranne_et_al_1996}, with the wavelength being calibrated using a Th-Ar lamp, which imprints emission lines over the wavelength domain of the spectrograph. 

Several other techniques have been demonstrated.
More recently, \citet{Anglada_2012} developed a new algorithm implemented in a piece of software called HARPS-TERRA to derive RVs from HARPS data using least square matching of each observed spectrum to a high signal-to-noise ratio (S/N) template derived from the same observation.
Besides the emission lamp technique, it is also common to use a gas absorption cell as wavelength calibrator. 
{A common absorption gas is iodine (I$_2$). Several spectrographs are equipped with this technology (e.g., the HIRES spectrograph at the Keck telescope).} Because the iodine absorption cell can be positioned directly in front of the slit, the spectrum entering the slit is, in fact, the product of the gas cell spectrum and the stellar one. This method allows a RV precion of $1 - 2\,\textrm{ms}^{-1}$ \citep{Marcy_1992, Butler_1996}.}

This level of precision is still not sufficient 
to find an Earth-like planet on a one-year orbit around a Sun-like star; for reference, the impact of Earth in the solar RV is  
of only $\sim 9\,\textrm{cms}^{-1}$. Therefore, a new generation of high-precision spectrographs is currently being planned with the objective of detecting Earth-mass planets; among these, ESPRESSO stands out. 
It aims for a precision of ten $\mathrm{cms}^{-1}$ \citep{Pepe_2010} and has, as one of the main science goals, detection of an earth-mass planet in the habitable zone of a Sun-like star. 
An increase in precision (and accuracy) implies a more detailed characterization of systematics and, in particular, 
the need to be more careful with contamination sources. % \citep[as e.g., see][]{Cunha_2013}.
{ As an example, the reader can look at the work of \citet{Cunha_2013}  which  presents a study of the impact by companions within the fiber in the RV calculation, where 
the contaminants are the secondary star of a binary system, a back/foreground star, and even just the moonlight.}

One {other} possible source of contamination is the Earth's atmosphere itself. Like all high-resolution spectrographs, HARPS is a ground-based spectrograph, 
and so, atmospheric spectra is imprinted on top of the stellar one.
To minimize the effect of this contamination on radial velocities computation, 3 of the 72 HARPS spectral orders, the ones in which the deeper telluric lines are present, 
are neglected in the RV computation \citep{Mayor_et_al_2003}. 
Moreover, in other spectral orders with deep telluric lines and/or strong blended lines, stellar masks used in the CCF
 take only a very few stellar lines into consideration  \citep{Baranne_et_al_1996, Pepe_2002}. 
But as the S/N of the spectra increases and the calibration methods are improved, the relative impact of 
shallow, not-neglected telluric lines, here designated as micro-telluric lines,  will become stronger.
Therefore, it becomes urgent to explore how to study the impact of these telluric lines of smaller depth and how to correct their effect
on the RV calculation.
 
The pursuit for the best method of correcting stellar spectra from atmospheric lines is not a recent matter.  
One possible way of doing it is to use standard stars, for which spectral features are well known, to determine 
the atmospheric spectrum \citep[see, e.g., ][]{Vacca_2003}.
Using simulations, \citet{bailey_2007} show that, for infrared observations, using an Earth atmospheric transmission spectrum modeled with SMART
(spectral mapping atmospheric radiative transfer model) leads to better correction 
of the telluric lines than using the ``division by a standard star'' technique.
Following the \citet{bailey_2007} work, \citet{Seifahrt_2010} developed a method of calibrating the wavelength
 of CRIRES spectrograph and correcting its observations from telluric lines 
by modeling the Earth atmospheric transmission  spectra  with the line-by-line radiative transfer model (LBLRTM). 
Also, \citet{Cotton_2013} used the division by modeled telluric spectra to remove of telluric features from Jupiter and Titan's infrared spectra, just to cite a few examples from an extensive list.  

The present paper follows the method that resorts to theoretical atmospheric transmission spectra 
to remove telluric features from stellar spectra. 
To achieve that, the atmospheric models,
LBLRTM and the Transmissions Atmosph\'eriques Personnalis\'ees Pour l'AStronomie (TAPAS), which are described in Sect \ref{TAPAS_LBLRTM}, are used.
In Sect. \ref{Impact} we present a study of the impact of the micro-telluric lines by adding an atmospheric spectrum to a 
synthetic Sun-like stellar spectrum, 
and in Sect. \ref {Correct} this effect is corrected by removing the atmospheric spectra from HARPS spectra.
We finish with a discussion about our work and present our conclusions in Sect. \ref{Discuss}.
}

\section{Atmospheric spectra with LBLRTM and TAPAS}\label{TAPAS_LBLRTM}
In this section we describe the two atmospheric models used in this work: LBLRTM and TAPAS, which are not independent.
The LBLRTM is an accurate model derived from the fast atmospheric signature code (FASCODE)
\citep {Clough_1981, Clough_1992}, which uses the HITRAN database 
(HIgh-resolution TRANsmission molecular absorption database) as the basis for the line parameters. 
This is a very versatile model, which besides the pre-defined models, also accepts a user-supplied atmospheric profile. 
In this work, for the LBLRTM, we used atmospheric data from the Air Resources Laboratory (ARL) 
-  Real-time Environmental Applications and Display sYstem (READY) Archived Meteorology \footnote{http://ready.arl.noaa.gov/READYamet.php}. 
Because ARL-READY only provides data up to 26 km of altitude, for higher atmospheric layers we used 
the MIPAS model atmospheres (2001) for the mid-latitude, nighttime \footnote{http://www.atm.ox.ac.uk/RFM/atm/}.
If one wants to have absolute control over the input parameters for the atmospheric spectra, LBLRTM is undoubtedly the model to use.  
The drawback in using LBLRTM is the lack of user friendliness in the input method  and the somewhat cryptic, highly technical user instructions. 
Therefore, in this work we also tested  a more user friendly model: TAPAS.

TAPAS  is a free online service that simulates atmospheric transmission \footnote{http://ether.ipsl.jussieu.fr/tapas/}. 
TAPAS makes use of the ETHER (Atmospheric Chemistry Data Centre) facility to interpolate within the ECMWF (European Centre for 
Medium-Range Weather Forecasts)
 pressure, temperature, and constituent profile at the location of the observing site and within six hours of the 
date of the observations, and it 
computes the atmospheric transmittance from the top of the atmosphere down to the observatory, 
based on the HITRAN and the LBLTRM code. TAPAS is much more user friendly than LBLRTM,
 although users have less control, having a much smaller number of input parameters. %, like the scanning functions.
For more details the user is referred to \citet{Bertaux_2014} .

\begin{figure*}[htb!]
   \centering 
  \includegraphics[width=0.99\textwidth ]{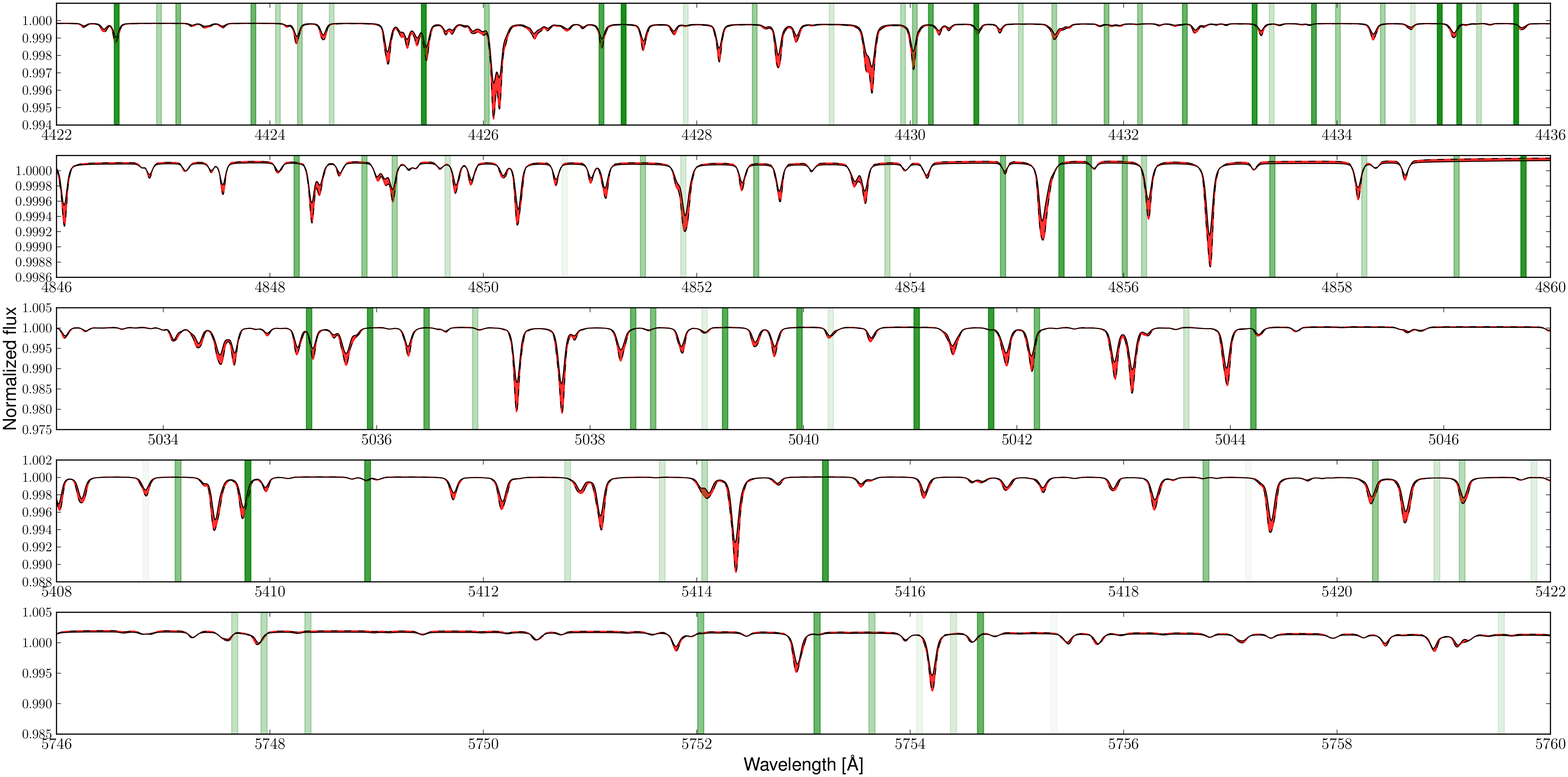}
  \caption{Normalized synthetic atmospheric spectra obtained with TAPAS (top black line) and LBLRTM (bottom black dashed line).
The vertical thick green lines represent the position, for a velocity $\mathrm{V=0}$, of stellar lines in the G2 stellar mask used by HARPS, and its intensity the weight in the CCF.}
  \label{fig:atm}
\end{figure*}

Figure \ref{fig:atm} shows the spectra obtained with both models, using identical physical conditions as input.
% The top black line represents the spectra obtained with TAPAS and the bottom dashed
% black line the LBLRTM model. 
 The two spectra overlap significantly but show a few differences (red shadow area between lines).  
% In Fig. \ref{fig:atm} the vertical green thick lines represent the position of "stellar lines" in the G2 mask used by HARPS pipeline, 
% where the  intensity corresponds to the weight of each line in the CCF. 
One can also see  in  Fig. \ref{fig:atm} that some telluric lines fall within or close to the mask lines, 
and they are taken into account in the RV calculation.

To compare the ability of the two models to reproduce the atmospheric transmission, a spectrum of a spectral type O5 star was corrected from 
the Earth's atmospheric absorption using both models.
A \textit{HD 46223} HARPS spectrum, taken during the program ID 185.D-0056(L),
 with a S/N of 220 at the center of the spectral order number 50 (which corresponds to a wavelength of $4372.8\,\AA$).
\begin{figure}[ht]
   \centering 
%\vspace{-15pt}
  \includegraphics[width=0.99\columnwidth ]{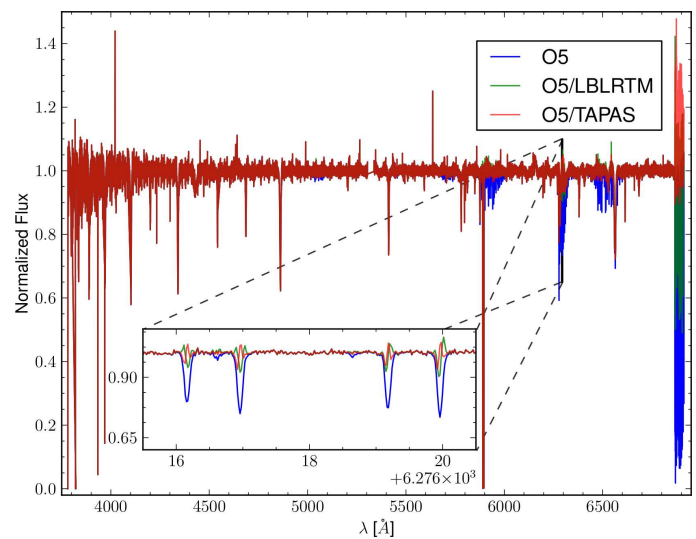}
  \caption{Blue: Normalized spectrum of the spectral type O5 \textit{HD 46223} star;
Green: Normalized spectrum of the spectral type O5 star \textit{HD 46223} divided by the atmospheric spectrum obtained with LBLRTM; 
Red:  Normalized spectrum of the spectral type O5 star \textit {HD 46223} divided by the atmospheric spectrum obtained with TAPAS.
}
  \label{fig:O6}
\end{figure}
The choice of an O5 star was motivated by the low number of spectral lines. 
Figure \ref{fig:O6} shows the normalized stellar spectra before correction, and after correction with LBLRTM  and with TAPAS.  
We also computed the standard deviations  $ \sigma_{O5/LBLRTM} $ and $ \sigma_{O5/TAPAS}$ in a part of the spectra without stellar 
lines, but with telluric spectral lines ($ 6474.6\,\AA< \lambda < 6518.3\,\AA $), as a measurement of the scattering of the corrected spectra.
 We obtained $\sigma$ values of 0.0078 and of 0.0081, respectively. 
Both models correct well the atmosphere, since both their standard deviations ($ \sigma_{O5/LBLRTM} $ and $ \sigma_{O5/TAPAS}$) are of the same order of magnitude, differing by a small amount ($\sim 3.7\%$.)  
% The differences between the two  models are small, and  both of them correct well the atmosphere, which is confirmed by the same order of magnitude
% of the  standard deviations, with a difference between them of only $3.7\%$. 
In this work we have chosen to use the TAPAS method because it is easier.

\section{Synthetic stellar spectra test}
\label  {Impact}
Before correcting HARPS spectra from telluric lines absorption, we first calculated the impact  of the  micro-telluric lines using synthetic 
stellar spectra of a solar star, i.e, the difference between the RV obtained
when using the stellar spectra with and without the atmosphere. 
The synthetic spectrum was built by extracting the spectral lines using VALD - Viena Atomic Line Database
\citep{Piskunov_1995,Ryabchikova_1997,Kupka_1999,Kupka_2000}, 
and then computing the spectrum using the running option \textit{synth} of MOOG \citep{Sneden_1973}. 
This spectra was multiplied by TAPAS atmospheric spectra for one night with H$_2$O vertical column of $14.52\, \mathrm{kgm^{-3}}$ . 
The water vapor content is relevant since the micro-telluric lines are partially H$_2$O, 
and thus the lines' depth varies with the water vapor content. Moreover, the lines' depth in the atmospheric spectra increases as the airmass increases.
 Therefore, we obtained TAPAS spectra for different values of the zenith angle (and so different values of airmass), using  the same time  of observation as input. 
In doing so we only consider zenithal angles of up to $60\,^\circ$, i.e, for airmass values below 2. 

For each simulated atmospheric spectra we then changed the RV of the star by shifting the stellar spectra
from $\mathrm  {RV} =-100 \,\mathrm{kms^{-1}}$ to $\mathrm  {RV} =100 \,\mathrm{kms^{-1}}$ in steps of $1\, \mathrm{kms^{-1}}$. 
The RV of each spectrum, with and without atmosphere, was calculated by cross-correlating it with a template \citep{Baranne_et_al_1996,Pepe_2002}. 
We then calculated the impact of the atmosphere as the difference between the RV of the original spectra (RV$_o$) and the RV of the modified spectra (RV$_m$):
\begin{equation}
\mathrm{Impact}  = RV_o - RV_m
\label{eq:impact}
\end{equation}

\begin{figure}[htb!]
   \centering 
  \includegraphics[width=0.99\columnwidth ]{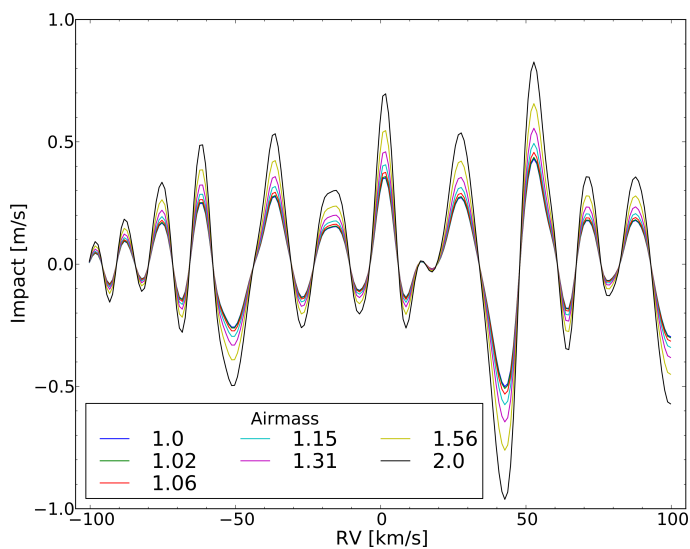}
  \caption{Impact of the atmosphere in the RV calculation for airmass values between 1 and 2. The inner line represents the airmass = 1
 and the outer the airmass = 2.}
  \label{synth}
\end{figure}

In the case of the synthetic stellar spectra, the original spectra will be the one without atmosphere, and the modified one
 the one with atmosphere. The results are presented in Fig. \ref{synth}, where one can see that the impact of the atmosphere will 
depend on the airmass and on the RV of the star, as expected. The impact of the different models scales with airmass.

\begin{figure}[htb!]
   \centering 
  \includegraphics[width=0.99\columnwidth ]{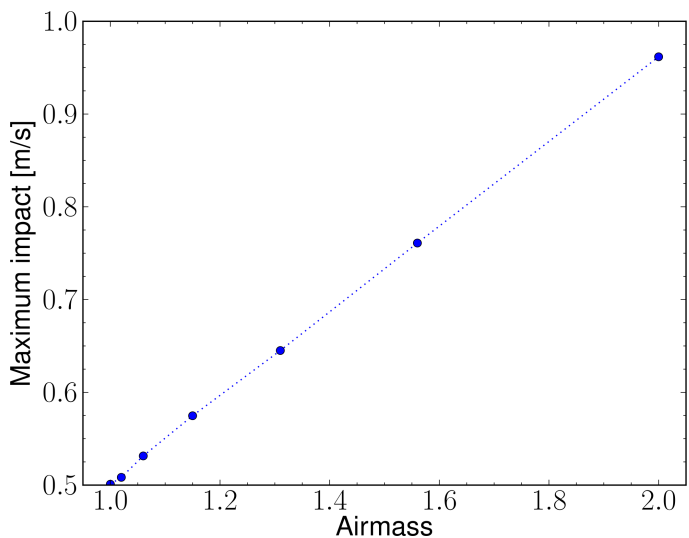}
  \caption{Maximum impact  for the  different airmass values. }
  \label{fig:maxImp}
\end{figure}

 When one compares the maximum impact for the  different values of airmass (Fig. \ref{fig:maxImp}), one can see that 
the micro-telluric lines can introduce an RV variation that is large enough to mimic or hide a planet. This is especially relevant when searching for small planets
 using observations on different nights with a large relative difference in the H$_2$O vertical column, or when the same star is observed several times in 
one night with large H$_2$O vertical column variation.    

\section{Correcting HARPS spectra from micro-telluric lines}
\label  {Correct}

\subsection{Data}\label{sec:data}

In his section we investigate the impact of the atmospheric micro-telluric lines 
on real spectra obtained with HARPS. 
To do so,  we  used \textit{18 Sco} time series of data obtained on the night of 2012-05-19 under the program ID 183.D-0729(B),
 allowing therefore the study of the impact of the airmass in the RVs along 
the night. The  S/N at the center of the spectral order number 50 (SN50) of the \textit{18 Sco} observations varies between 124 and 212.4,
 and the H$_2$O vertical column predicted by TAPAS varies between 3.83 and 4.51$\,\mathrm{kg\,\textrm{m}^{-3}}$.

To complement the analysis of the \textit{18 Sco} data, we also used spectra of three other stars, ranging in spectral type from G to M
- Tau Ceti (G8), HD85512 (K5), and Gl436 (M1).
As in our previous work on the impact of stellar companions on precise radial velocities \citep{Cunha_2013}, where these stars were also used, 
we chose these stars because they were the ones for which we could find HARPS spectra for their spectral type with the highest S/N.
We used spectra with SN50 ranging  between 30 for the M star, and 439 for the G star. 
The values of S/N of the used spectra can be found in the table of Appendix \ref{tbl:sn50}, 
along with the airmass and the Program ID for each observation.

\subsection{Method}
\label{method}
To determine the impact of the atmosphere in the RV calculation using HARPS spectra, we divided the HARPS stellar
 spectra by the normalized TAPAS synthetic atmospheric spectra. The TAPAS spectra used for  each stellar spectra correction from the atmosphere
is obtained using, as input parameters: the Observatory (ESO La Silla Chile); the exact date and hour of the observation; 
the spectral range in wavenumber units ($14450 - 26550\: \mathrm{cm}^{-1}$); the instrumental function as Gaussian; 
the ARLETTY atmospheric model{, which is an ETHER atmospheric  profile computed by using the nearest in time of the  ECMWF meteorologic field observation};
 a resolution power of 115000 (HARPS' resolution); a sampling ratio{, i. e., the number of points on which the convolved transmission 
will be sampled for each interval of FWHM,} of 10; and the Zenithal angle of the star.
 
All the remaining preferences of the TAPAS request form were the default ones.
To convert the wavenumber (WVNR) to the air wavelength (AIR) used in HARPS spectra, we first converted into vacuum wavelength (VAC):
\begin{equation}\label{eq:wnr-vac}
 WVNR\left[\mathrm{cm}^{-1}\right]=\frac{10^8}{VAC\left[\AA \right]}.
\end{equation}
%where  \textit{WVNR} is the wavenumber and \textit{VAC} is the vaccum wavelength. 
Then we used the IAU standard conversion from AIR to VAC as given by \citet{Morton_1991}
\begin{equation}\label{eq:air-vac}
 AIR \left[\AA\right]=\frac{VAC\left[\AA\right]}{(1.0 + 2.735182\times10^{-4}+\frac{131.4182}{VAC\left[\AA\right]^2}+\frac{2.76249\times10^8}{VAC\left[\AA\right]^4}}.
\end{equation}
 %where \textit{AIR} is the air wavelength. 

Because the atmospheric spectra wavelength grid is different than HARPS wavelength grid, we interpolated
 the atmospheric wavelength grid to the HARPS grid. 
On top of that, we shifted the atmospheric spectra  of $\pm 1500\,\mathrm{ms}^{-1}$ in steps of $1\,\mathrm{ms}^{-1}$, 
 we divided the normalized stellar spectra by each shifted atmospheric spectra, and we calculate which shifted spectra minimizes the difference
between the stellar spectrum before and after atmospheric correction by minimizing the $\chi^2$.
{This will be the atmospheric spectra that best corrects for the atmosphere, since more telluric lines of the atmospheric spectrum will coincide with the telluric lines present in the stellar spectrum. 

The $\pm 1500\,\mathrm{ms}^{-1}$ value was chosen to make certain that the shift between  the telluric lines of the atmospheric spectrum and the telluric  lines  in the stellar one, either because of the  interpolation itself or because of the RV variations of the  telluric lines \citep{Figueira_2010b, Figueira_2012}, is within  that range.
}
Then we ran the HARPS pipeline to determine the RV of the stars both when considering the atmosphere and after removing it.
Recalling Eq. \ref{eq:impact}, in this case the impact is calculated considering  
RV$_o$ as the RV calculated using the original HARPS spectra with the atmosphere, 
and RV$_m$ as the RV of the HARPS spectra after removing the atmosphere.

\subsection{Results}
\label{results}
Following the work done in Sect. \ref{Impact}, we wanted to investigate the impact of micro-telluric lines, and its dependence on the airmass, now using HARPS 
spectra. We  used \textit{18 Sco} observations taken in the same night, permitting us therefore to have data from the same star during the same 
night and with different airmass values. 
\begin{figure}[htb!]
   \centering 
  \includegraphics[width=0.99\columnwidth ]{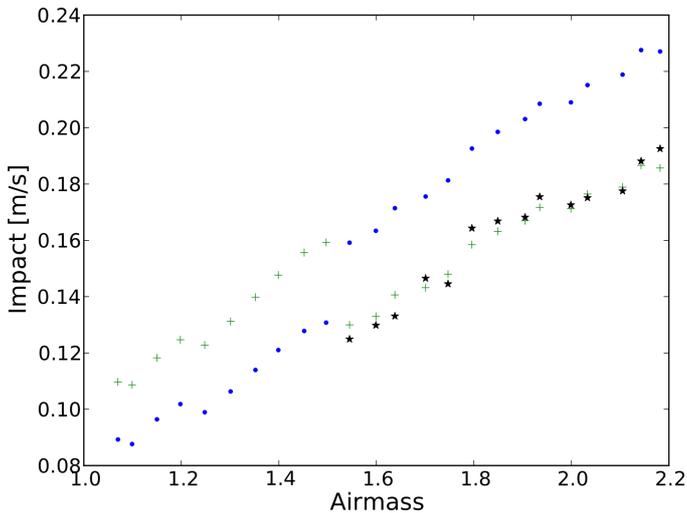}
  \caption{Impact of the atmosphere in the RVs for 18 Sco for different values of airmass. All the spectra were taken during the night of 2012/05/19. 
The blue dots represent the impact calculated using the default TAPAS spectra, the green crosses ($+$) in the left part of the plot
represent the impact using the TAPAS spectra for
the time corresponding to spectra with airmass 1.55, and green $+$ crosses in the right part the impact using the TAPAS spectra for
the hour corresponding to spectra with airmass 1.5. 
The black stars represent the impact using adjusted H$_2$O transmission spectra. }
  \label{fig:imp_18sco}
\end{figure}
If one considers the blue dots in Fig. \ref{fig:imp_18sco}, one can see the impact of the atmosphere depends on the airmass.
 One can  easily see that the impact seems to be linearly proportional to the airmass, 
except for airmasses around 1.5,
where one can observe a jump in the impact value. 
This jump might be because the data used in the TAPAS atmospheric model is obtained only every six hours. 
Thus, although one of TAPAS input parameters is the exact hour of the observation, the atmospheric output will  be based on an approximated hour, 
and for one full night of observations, one will have at most three different modeled atmospheric spectra. 

To test that the jump in the impact around $\mathrm{airmass} = 1.5$ was due to a change in the atmospheric data, we repeated our calculation, 
 now using  TAPAS spectra with a wrong time of observation:
we used the a TAPAS spectra with the time stamp of the stellar spectrum with $\mathrm{airmass} = 1.55$ 
for the correction of  the stellar spectra with $\mathrm{airmass}\leq1.5$
(left part of Fig. \ref{fig:imp_18sco} ), and  TAPAS spectra with the time stamp corresponding to the
 spectrum  with airmass of 1.5 for the correction of the stellar spectra with $\mathrm{airmass}\geq1.5$ 
 (right part of Fig. \ref{fig:imp_18sco}). 
By doing so we end up with two nearly parallel lines for the impact. Each line corresponds to a correction with a different modeled atmospheric spectrum.

This result corroborates the hypothesis that the original jump in the impact was due to using two different atmospheric data sets.
Moreover, we visually checked that the atmospheric lines were being corrected well, and we found that for airmasses greater that 1.5, telluric lines were being 
over corrected. Thus, for each of these spectra, we adjusted the H$_2$O transmission by using the power law $T^X(H_2O)$, 
T being the transmission and X the adjusting factor \citep{Bertaux_2014}. To choose the X value that corrected the H$_2$O transmission better, we chose a small 
part of the spectrum with no stellar lines ($\lambda = 6483.25 - 6490.25 \AA$), and we minimized the standard deviation for this corrected part of the spectrum.   
The black stars in Fig. \ref{fig:imp_18sco} show the calculated impact in the radial velocities, when used for telluric correction the adjusted transmission. 
We find a difference between the impact calculated with and without adjusted transmission that can go up to  $4\,\mathrm{cms}^{-1}$,
 which is close to the magnitude of the error of the impact calculation (see Sect. \ref{Discuss}).

\begin{figure}[htb!]
   \centering 
  \includegraphics[width=0.99\columnwidth ]{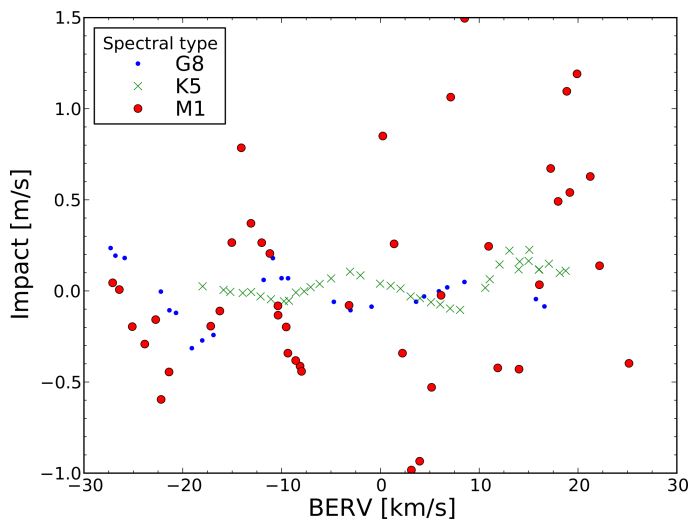}
  \caption{Impact of the atmosphere in the radial velocities for different BERV (barycentric Earth radial velocity) values. 
A star of spectral type G is represented in small blue dots, a K star in green x crosses, and an M in big red dots.  }
  \label{fig:imp_stype}
\end{figure}

Besides studying the effect of the atmosphere on the RV calculation in the night, 
we also investigated whether the impact of the micro-tellurics would vary along the year for stars of spectral type G8, K5, and M1. 
For the CCF of the each spectrum, we used the corresponding DRS stellar mask, i.e., for the G8, K5, and M1 stars we used the G2, K5, and M2 masks, respectively. 
The result is shown in Fig. \ref{fig:imp_stype}, where one can see the variation in impact of the atmosphere in the RVs calculations
 with the barycentric Earth radial velocity (BERV) for stars of spectral type G, K, and M.
As one has can see in Fig. \ref{synth}, telluric lines have different impacts depending on the RV of the star. Thus, as the RV of a star changes along the year with respect to the Earth, 
the impact of the atmosphere also varies as the Earth goes around the Sun, and as the BERV varies.
 The maximum absolute values of the Impact for these G,K, and M stars is of 31.4, 22.6, and 149.6 $\mathrm{cms}^{-1}$, respectively. 
We also calculated the standard deviation ($\sigma$) for the relations between the impact and the stellar RV of the G2 synthetic spectra (Fig. \ref{synth})
 and between the impact and the BERV (Fig. \ref{fig:imp_stype}) for the different spectral types.
For the impact G2 star, we present $\sigma$ for an airmass of 1.0, 1.15, and 2.0. For the stellar spectra obtained with HARPS, we calculated $\sigma$
 considering all the impacts of each star, and also neglecting impacts obtained with airmasses above 1.15 for the G8 and K5 stars and above 2.0 for the M1 star.
These values are presented in Table \ref{tbl:std}. One can see that $\sigma$ increases since the observations are done with a higher airmass.
By considering only observations made with the star at an airmass of less than 1.15, we obtained a difference in $\sigma$ of 15.1\% for the G8 star and of 2.4\% for the K5, and 
neglecting airmasses above 2.0 for the M1 star, we found a difference of 1.6\%.
\begin{table}
\caption{Standard deviation, $\sigma$, of the relation between the impact of the atmosphere and the radial velocity of the star,
 depending on the airmass at the time of the observation.}             % title of Table
\label{tbl:std}      % is used to refer this table in the text
\centering                          % used for centering table
\begin{tabular}{c c c }        % centered columns (4 columns)
\hline\hline                 % inserts double horizontal lines
\textbf{Spect. type} & \textbf{Airmass} & $\bf{\sigma}$  \\    % table heading 
		    &  			 & $\bf{[\mathrm{ms}^{-1}]}$  \\
\hline                        % inserts single horizontal line
G2 (synth.) & 1.0  & 0.173  \\      % inserting body of the table
            & 1.15 & 0.199  \\
            & 2.0  & 0.336  \\ \hline
G8          &  all & 0.139  \\
            &$<1.15$& 0.118  \\\hline
K5          &  all & 0.085  \\
            &$<1.15$  & 0.83  \\\hline
M1          &  all & 0.555  \\
            &$<2.0 $  & 0.546  \\\hline
\hline                                   %inserts single line
\end{tabular}
\end{table}
One should note that the spectra used to calculate the impact for the different spectral types was taken on different nights and at different hours. 
Therefore, it is difficult to make a direct comparison between the impact of the atmosphere for the different spectral types. 
Nevertheless, we can envision a periodic variation in the impact with the BERV, which is clearer for the G8 and K5 stars. 

\section{Discussion and Conclusions}
\label{Discuss}

With our work we showed that micro-telluric lines should be considered in the calculation of RVs at the level of sub-$\mathrm{m}\,\mathrm{s}^{-1}$. 
Our test case using synthetic spectra (see Fig. \ref{fig:maxImp}) shows a maximum impact of $96.2 \,\mathrm{cms}^{-1}$ for the case in which we have an airmass of 2.
This is $46\,\mathrm{cms}^{-1}$ more than when the star is at its zenith ($\mathrm{airmass} = 1$. 
Thus, even if the $50.1\,\mathrm{cms}^{-1}$ of impact for the minimum value of the airmass  could be explained  by systematics, the difference of  $46\,\mathrm{cms}^{-1}$
can only be explained as a consequence of the micro-telluric lines. 
A  $46\,\mathrm{cms}^{-1}$ impact is whithin the limit of HARPS precision, and is greater than the predicted precision 
for ESPRESSO. Thus the Earth's atmospheric absorption should be corrected for when looking for Earth-size exoplanets.

In Sect. \ref{Correct} we divided HARPS spectra by atmospheric spectra modeled with TAPAS  to correct the effect of the micro-telluric lines. 
When we corrected stellar spectra of one star taken during one night (Fig. \ref{fig:imp_18sco}),
 we obtained a  linear variation of the impact with the airmass, as observed with the synthetic spectra.
However, the atmospheric data is only updated every six hours.
Therefore the atmospheric spectra used in the correction might not be the spectra corresponding to the exact hour of the stellar observation,
and so it may lead to an under-overcorrection of the stellar spectra. This is especially true  when there is a change in the weather conditions. 
This should be the reason for the observed jump of $2.8\,\mathrm{cms}^{-1}$ in the impact, previously presented  in Fig. \ref{fig:imp_18sco}.

The changes in the weather conditions and the airmass variation are not the only factors involved in the micro-telluric effect. 
As seen in Fig. \ref{fig:imp_stype}, there is also a variation with BERV, i.e., a periodic variation during the year. 
This periodic variation is clear, particularly for the G8 and K5 stars. For the M1 star this variation is not so clear. 
This might be because of the other factors that contribute to the atmospheric impact in the RVs: the spectra were taken during the year, so 
they were taken on different nights. The ESO atmospheric-conditions archive for La Silla is not always available for the nights of observation,
 but the most probable scenario is the one in which there are significantly different  weather conditions for the several nights.
 Also, the airmass is not constant, as one can see from Table \ref{tbl:sn50}. 
 Therefore, in addition to the BERV effect, one also have the effect of the other factors that contribute to the impact of the micro-telluric lines on the RV calculation.
 From Fig. \ref{fig:imp_stype} one can also see that the impact of Earth's atmosphere on RV probably seems more problematic for the M1 star. This can be because of
 the high airmass values. This star is always near the horizon, so its airmass values are always above 1.7.
 {One way of testing our micro-telluric correction is to calculate the \textit{rms} for the RV points taken in one night, before and after atmospheric correction. The \textit{rms} is expected  to be smaller after the correction of the atmospheric lines; i.e., the RV points will be less dispersed.  Unfortunately, currently available HARPS data do not permit this to be confirmed. We would need asteroseismology time series observation for stars of spectral type K-M, which we expect to suffer a higher contamination from the atmosphere (Fig. \ref{fig:imp_stype}). But there is no asteroseismology data for stars of these spectral types. We tested then for stars of spectral type G: \textit{18 Sco} (G2) and \textit{Tau Ceti} (G8). {Although for some data sets there was, as expected, a decrease in the dispersion, the maximum improvement was marginal: $0.86\,\mathrm{cms}^{-1}$ for  \textit{18 Sco} and $0.26\,\mathrm{cms}^{-1}$ for \textit{Tau Ceti}. With current data we thus cannot completely test the gain when correcting the telluric absorption.}
 The impact of the micro-telluric lines in these G-type stars RV is in the range $10-20\,\mathrm{cms}^{-1}$, which is lower than the value of HARPS uncertainty along one night \citep{Dumusque_2011a}. Therefore, it is most probably dominated by instrumentation/calibration noise and not by the atmospheric contamination.} 
 
In addition to the factors considered in this work, others should be investigated. 
Of these we highlight the wind direction and intensity variations, which may induce a shift the atmospheric lines.  

We also want to note the lack of error bars in our figures when working with HARPS spectra.
Because the impact value is estimated  using the same original stellar spectrum with and without atmosphere  
and the removal of the atmospheric lines is made using an error-free synthetic atmospheric spectrum,  
the error of the impact will just be the difference between the estimated RV uncertainties for the case with and without atmosphere. 
To have an idea of these errors, we present the \textit{18 Sco} values when the star is at its lowest and its highest airmass. 
In the first case ($\mathrm{airmass} = 1.070 $), the error will be of $2.1\,\mathrm{cms}^{-1}$, 
and in the second ($\mathrm{airmass} =2.182 $) it will be $3.4\,\mathrm{cms}^{-1}$.

\begin{acknowledgements}
{We acknowledge the support from Funda\c{c}\~ao para a Ci\^encia e a Tecnologia (FCT, Portugal)
through FEDER funds in program COMPETE, as well as through national funds, in the form of grants
 PTDC/CTE-AST/120251/2010 (COMPETE reference FCOMP-01-0124-FEDER-019884), RECI/FIS-AST/0176/2012
(FCOMP-01-0124-FEDER-027493), and RECI/FIS-AST/0163/2012 (FCOMP-01-0124-FEDER-027492).
We also acknowledge the support from the European Research Council/European Community under the FP7
through Starting Grant agreement number 239953.
NCS and PF were supported by FCT through the Investigador FCT
contract reference IF/00169/2012 and IF/01037/2013 and POPH/FSE (EC) by FEDER funding through the program
"Programa Operacional de Factores de Competitividade - COMPETE. }

\end{acknowledgements}

\bibliographystyle{aa}
\bibliography{Diana}
\newpage
\onecolumn
\appendix 
\section{Star properties}\label{tbl:sn50}
\begin{center}
\begin{longtable}{c c c c c}
\caption{Star properties in our simulations.}\\
\hline
\textbf{Object} & \textbf{Spec. type} & \textbf{SN50}\footnotemark & \textbf{Airmass}& \textbf{Prog. ID}\\ \hline\hline
\endfirsthead
\multicolumn{5}{c}%
{{\bfseries \tablename\ \thetable{} -- continued from previous page}} \\
\hline
\textbf{Object} & \textbf{Spec. Type}& \textbf{SN50} & \textbf{Airmass} & \textbf{Prog. ID}\\ \hline\hline
\endhead
\hline \multicolumn{5}{|r|}{{Continued on next page}} \\ \hline
\endfoot
\hline \hline
\endlastfoot
   {Tau Ceti} &{G8}        &    286 &    1.208 &   082.C-0315(A)\\
         &   &    229 &    1.290 &      082.C-0315(A)\\
         &   &    164 &    1.042 &      083.C-1001(A)\\
         &   &    196 &    1.039 &      083.C-1001(A)\\
         &   &    308 &    1.039 &      083.C-1001(A)\\
         &   &    374 &    1.034 &      083.C-1001(A)\\
         &   &    196 &    1.028 &      083.C-1001(A)\\
         &   &    287 &    1.079 &      083.C-1001(A)\\
         &   &    255 &    1.033 &      083.C-1001(A)\\
         &   &    273 &    1.033 &      084.C-0229(A)\\
         &   &    269 &    1.195 &      084.C-0229(A)\\
         &   &    439 &    1.030 &      084.C-0229(A)\\
         &   &    381 &    1.347 &      084.C-0229(A)\\
         &   &    346 &    1.396 &      084.C-0229(A)\\
         &   &    386 &    1.401 &      084.C-0229(A)\\
         &   &    409 &    1.029 &      084.C-0229(A)\\
         &   &    198 &    1.028 &      084.C-0229(A)\\
         &   &    193 &    1.123 &      084.C-0229(A)\\
         &   &    245 &    1.129 &      084.C-0229(A)\\
         &   &    218 &    1.090 &      084.C-0229(A)\\
         &   &    236 &    1.067 &      084.C-0229(A)\\
         &   &    274 &    1.067 &      084.C-0229(A)\\
         &   &    264 &    1.066 &      084.C-0229(A)\\
         &   &    300 &    1.067 &      084.C-0229(A)\\\hline
{HD85512} & {K5}  &    201 &    1.095 &    072.C-0488(E)\\
 & &    248 &    1.032 &     072.C-0488(E)\\ 
 & &    191 &    1.032 &     072.C-0488(E)\\ 
 & &    184 &    1.047 &     072.C-0488(E)\\ 
 & &    163 &    1.037 &     072.C-0488(E)\\ 
 & &    177 &    1.032 &     082.C-0315(A)\\ 
 & &    187 &    1.132 &     082.C-0315(A)\\ 
 & &    169 &    1.032 &     082.C-0315(A)\\ 
 & &    145 &    1.056 &     082.C-0315(A)\\ 
 & &    160 &    1.379 &     082.C-0315(A)\\  
 & &    195 &    1.140 &     082.C-0315(A)\\ 
 & &    188 &    1.106 &     082.C-0315(A)\\ 
 & &    205 &    1.058 &     082.C-0315(A)\\ 
 & &    206 &    1.064 &     084.C-0229(A)\\ 
 & &    172 &    1.032 &     086.C-0230(A)\\ 
 & &    178 &    1.115 &     086.C-0230(A)\\ 
 & &    162 &    1.267 &     086.C-0230(A)\\ 
 & &    181 &    1.035 &     086.C-0230(A)\\ 
 & &    196 &    1.043 &     086.C-0230(A)\\ 
 & &    136 &    1.033 &     086.C-0230(A)\\ 
 & &    143 &    1.032 &     086.C-0230(A)\\ 
 & &    152 &    1.067 &     086.C-0230(A)\\ 
 & &    185 &    1.047 &     086.C-0230(A)\\ 
 & &    180 &    1.033 &     087.C-0990(A)\\ 
 & &    202 &    1.038 &     087.C-0990(A)\\ 
 & &    128 &    1.126 &     087.C-0990(A)\\ 
 & &    132 &    1.073 &     087.C-0990(A)\\ 
 & &    159 &    1.034 &     087.C-0990(A)\\ 
 & &    197 &    1.076 &     087.C-0990(A)\\ 
 & &    167 &    1.134 &     087.C-0990(A)\\ 
 & &    182 &    1.036 &     087.C-0990(A)\\ 
 & &    161 &    1.078 &     087.C-0990(A)\\ 
 & &    167 &    1.122 &     087.C-0990(A)\\ 
 & &    183 &    1.059 &     087.C-0990(A)\\ 
 & &    195 &    1.398 &     088.C-0011(A)\\ 
 & &    170 &    1.066 &     088.C-0011(A)\\ 
 & &    134 &    1.069 &     088.C-0011(A)\\ 
 & &    163 &    1.204 &     088.C-0011(A)\\ 
 & &    149 &    1.287 &     088.C-0011(A)\\ 
 & &    155 &    1.377 &     088.C-0011(A)\\ 
 & &    137 &    1.152 &     088.C-0011(A)\\ 
 & &    152 &    1.210 &     088.C-0011(A)\\\hline
{Gl436} & {M1} &     34 &    1.788 &   072.C-0488(E)\\ 
 & &     48 &    1.792 &      072.C-0488(E)\\ 
 & &     44 &    1.829 &      072.C-0488(E)\\ 
 & &     48 &    1.790 &      072.C-0488(E)\\ 
 & &     47 &    1.785 &      072.C-0488(E)\\ 
 & &     39 &    1.788 &      072.C-0488(E)\\ 
 & &     48 &    1.815 &      072.C-0488(E)\\ 
 & &     41 &    1.801 &      072.C-0488(E)\\ 
 & &     30 &    1.897 &      072.C-0488(E)\\ 
 & &     43 &    1.807 &      072.C-0488(E)\\ 
 & &     32 &    1.781 &      072.C-0488(E)\\ 
 & &     54 &    1.781 &      072.C-0488(E)\\ 
 & &     44 &    2.256 &      072.C-0488(E)\\ 
 & &     39 &    2.046 &      072.C-0488(E)\\ 
 & &     43 &    2.164 &      072.C-0488(E)\\ 
 & &     45 &    1.780 &      072.C-0488(E)\\ 
 & &     45 &    1.780 &      072.C-0488(E)\\ 
 & &     41 &    1.781 &      072.C-0488(E)\\ 
 & &     35 &    1.780 &      072.C-0488(E)\\
 & &     30 &    2.098 &      072.C-0488(E)\\ 
 & &     50 &    1.831 &      072.C-0488(E)\\ 
 & &     50 &    1.785 &      072.C-0488(E)\\ 
 & &     47 &    1.825 &      072.C-0488(E)\\ 
 & &     44 &    1.843 &      072.C-0488(E)\\ 
 & &     51 &    1.781 &      072.C-0488(E)\\ 
 & &     34 &    1.781 &      072.C-0488(E)\\ 
 & &     45 &    1.821 &      072.C-0488(E)\\ 
 & &     63 &    1.788 &      072.C-0488(E)\\ 
 & &     35 &    1.785 &      072.C-0488(E)\\ 
 & &     42 &    1.784 &      072.C-0488(E)\\ 
 & &     51 &    2.035 &      072.C-0488(E)\\ 
 & &     47 &    1.781 &      072.C-0488(E)\\ 
 & &     44 &    1.791 &      072.C-0488(E)\\ 
 & &     35 &    1.860 &      072.C-0488(E)\\ 
 & &     32 &    1.841 &      082.C-0718(B)\\ 
 & &     46 &    1.780 &      082.C-0718(B)\\ 
 & &     42 &    1.782 &      082.C-0718(B)\\ 
 & &     47 &    1.798 &      082.C-0718(B)\\ 
 & &     48 &    1.806 &      082.C-0718(B)\\ 
 & &     46 &    1.781 &      082.C-0718(B)\\ 
 & &     45 &    1.783 &      082.C-0718(B)\\ 
 & &     42 &    1.901 &      082.C-0718(B)\\ 
 & &     44 &    1.836 &      082.C-0718(B)\\ 
 %\end{tabular}
\footnotetext{S/N at the center of the spectral order number 50, corresponding to a wavelength of $4372.8\,\AA$.}
 %\end{center}
% \label{tbl:sn50}
 \end{longtable}
 \end{center}

\end{document}